\documentstyle[12pt]{article}
\newcommand{\Eq}[1]{Eq.\,(\ref{#1})}
\def\beq{\begin{equation}}
\def\eeq{\end{equation}}
\def\bear{\begin{eqnarray}}
\def\ear{\end{eqnarray}}
\newcommand{\F}{F_a}
\newcommand{\T}{\theta}

\newcommand{\Ref}[1]{Ref.\,\cite{#1}}
\newcommand{\Refs}[1]{Refs.\,\cite{#1}}

\author{ {\large M.Yu.~Khlopov}{\small $^{1,2,3}$}\thanks{e-mail: mkhlopov@orc.ru}
 {\large ,~A.S.~Sakharov}{\small $^{1,3}$}\thanks{e-mail: sakhas@landau.ac.ru}
  {\large ~and~D.D.~Sokoloff}{\small $^4$}\thanks{e-mail: sokoloff@dds.srcc.msu.su}}
\title{{\LARGE{\bf The large--scale modulation of the density distribution 
 in standard axionic CDM and its cosmological and physical impact}}
\footnote{Talk presented at Workshop on 
Fundamental Physics at the Birth of the 
Universe II, Roma, May 19--24, 1997}}
\date{{\small{\it $^1$Center for CosmoParticle Physics "Cosmion"\\ 
$^2$Institute of Applied Mathematics,
Miusskaya Pl.4, 125047 Moscow, Russia\\ 
$^3$Moscow Engineering Physics Institute 
(Technical University),  Kashirskoe Sh.31, 115409 Moscow, Russia\\
$^4$Department of Physics, 
Moscow State University, 119899 Moscow, Russia}}}

\begin{document}
\maketitle

\begin{abstract}
  	It is shown, that the energy density of coherent axion field oscillations 
in 
the cosmology of standard invisible axion should be distributed in the Universe
 in the form of archioles, being nonlinear inhomogeneous structure, reflecting 
the large scale distribution of Brownian structure of axion strings in the very 
early Universe. Spectrum of inhomogeneities, generated by archioles, is obtained 
and their effects in the spectrum and quadrupole anisotropy of relic radiation 
are 
considered. The axionic--string--decay--model--independent restriction on the scale 
of 
axion interaction is obtained.
\end{abstract}

\section{Introduction}
The modern theory of large-scale structure of the Universe is based on the 
assumption 
that this structure is formed as the result of development of gravitational 
instability 
from small initial perturbations of density or gravitational potential. As a 
rule, these 
perturbations are gaussian, but some version of nongaussian perturbations have 
also 
been considered.
 
In this paper we analyze the problem, inherent to practically all the 
cosmological cold 
dark matter models of invisible axion, that concerns primordial inhomogeneity in 
the 
distribution of the energy of coherent oscillations of the axion field. This 
problem, 
referred to as the problem of {\it archioles}, invokes nongaussian component in 
the 
initial perturbations for axionic cold dark matter. Archioles are the formation 
that 
represents a replica of the percolation Brownian vacuum structure of axionic 
walls 
bounded by strings, which is fixed in the strongly inhomogeneous primeval 
distribution of cold dark matter. They can give rise to interesting alternative 
scenarios 
of structure formation that relate the mechanism responsible for structure 
formation to 
inhomogeneities of the type of topological defects. 

The analysis of observable effects associated with {\it archioles} leads to a 
new 
model-independent constraint on the mass of invisible axion. Such analysis is 
also 
very useful for further development of full cosmological theories, based on the 
model 
of horizontal unification (MHU), which has been proposed in \Ref{13} as the 
minimal phenomenology of everything, including the physics of inflation, 
bariosynthesis, and dark matter. In particular, the combination of archioles 
effect with 
the consideration of nonthermal symmetry restoration in the horizontal phase 
transitions on inflation stage, puts the upper limit on the scale of family 
symmetry 
breaking (the main parameter of MHU) and consequently severely reduces the set 
of 
possible realizations of dark matter scenarios in this model.

\section{Formation of the archiole structure in the early 
\mbox{Universe}}
In the standard invisible axion scenario \Ref{1} the breaking of the Peccei-
Quinn 
symmetry is induced by the complex $SU(3)\bigotimes SU(2)\bigotimes U(1)$ -- 
singlet Higgs field $\phi$ with a "Mexican hat" potential
\beq
\label{Vmex}
V(\phi )=\frac{\lambda}{2}\left(\phi^+\phi -\F^2\right)^2
\eeq
Such field can be represented as $\phi =\F\exp (i\T )$, where $\theta =a/\F$ and 
$a$ 
is the angular Goldstone mode-axion. QCD instanton effects remove the vacuum 
degeneracy and induce effective potential for $\T$
\beq
\label{T}
V(\T )=\Lambda^4_1(1-\cos(\T N))
\eeq
Below we will simply assume for standard axion that $N=1$. In the context of 
standard big bang scenario it is usually assumed that the phase transition with 
$U(1)$ --
 symmetry breaking occurs when the Universe cools below the temperature 
$T\cong\F$. Thus, in the standard case the crucial assumption is that from the 
moment of the PQ phase transition and all the way down to the temperatures 
$T\cong\Lambda_{QCD}$, the bottom of the potential \Eq{Vmex} is exactly flat and 
there is no preferred value of $a$ during this 
period (the term given by \Eq{T} vanishes). Consequently, 
at the moment of the QCD phase transition, when the instanton effects remove 
vacuum degeneracy, $a$ rolls to the minimum and starts coherent oscillations 
(CO) 
about it with energy density \Ref{1}
\beq
\label{dens}
\rho_a(T,\T)=19.57\left(\frac{T^2_1m_a}{M_P}\right)\left(\frac{T}{T_1}\right)^3\
T^
2\F^2 
\eeq
The coherent axion field oscillations turn on at the moment $\tilde t\approx 
8.8\cdot 
10^{-7}$s.
 
It is generally assumed, that PQ transition takes place after inflation and the 
axion 
field starts oscillations with different phase in each region causally connected 
at 
$T\cong\F$, so one has the average over all the values to obtain the modern
axion density. Thus in the standard cosmology of invisible axion, it is usually 
assumed that the energy density of coherent oscillations is distributed 
uniformly and 
that it corresponds to the averaged phase value of $\bar{\T} =1$ 
($\bar{\rho_a}=\rho 
(\bar{\T})$). However, the local value of the energy density of coherent 
oscillations 
depends on the local phase $\T$ that determines the local amplitude of these 
coherent 
oscillations. It was first found in \Ref{2}, that the initial large-scale (LS) 
inhomogeneity of the distribution of $\T$ must be reflected in the distribution 
of the 
energy density of coherent oscillations of the axion field. Such LS modulation 
of the 
distribution of the phase $\T$ and consequently of the energy density of CO 
appears 
when we take into account the vacuum structures leading to the system of axion 
topological defects. 

As soon as the temperature of Universe becomes less then $\F$, the field $\phi$ 
acquires the vacuum expectation value (VEV) $\langle\phi\rangle =\F\exp {(i\T 
)}$, 
where $\T$ varies smoothly at the scale $\F^{-1}$. The existence of 
noncontractable 
closed loops that change the phase by $2\pi n$ leads to emergence of axion 
strings. 
These strings can be infinite or closed. The numerical simulation of global 
string 
formation \Ref{3} revealed that about 80\% of the length of strings corresponds 
to 
infinite Brownian lines. The remaining 20\% of this length is contributed by 
closed 
loops. Infinite strings form a random Brownian network with the step 
$L(t)\approx t$. 
After string formation when the temperature becomes as low as 
$T\approx\Lambda_{QCD}$, the \Eq{T} makes a significant contribution to the 
total 
potential so that the minimum of energy corresponds to a vacuum with $\T =2\pi 
k$, 
where $k$ is an integer -- for example, $k=0$. However, the vacuum value of the 
$\T$ 
cannot be zero everywhere, since the phase must change by $\Delta\T =2\pi$ upon 
a 
loop around a string. Hence, we come from the vacuum with $\T =0$ to the vacuum 
with $\T =2\pi$  as the result of such circumvention. The vacuum value of $\T$ 
is 
fixed at all points with the exception of the point $\T =\pi$. At this point, a 
transition 
from one vacuum to another occurs, and the vacuum axion wall is formed 
simultaneously with CO turning on. The width of such wall bounded by string is 
$\delta\cong m_a^{-1}$. Thus, the initial value of  $\T$ must be close to $\pi$ 
near 
the wall, and the amplitude of CO in \Eq{dens} is determined by the difference 
of the 
initial local phase $\T (x)$ and the vacuum value, which is different from the 
one
of the true vacuum only in 
a narrow region within the wall of thickness $\delta\cong m_a^{-1}$. Therefore 
in 
this region we can write $\T (x)=\pi -\varepsilon (x)$, where \Ref{2} 
$\varepsilon 
(x)=2\tan^{-1}(\exp (m_ax))$ and $x\cong m_a^{-1}$. Thereby the energy density 
of 
CO in such regions is given by 
\beq
\label{adens}
\rho^A\approx\pi^2\bar{\rho_a}
\eeq
And so we obtain, that the distribution of CO of axion field is modulated by 
nonlinear 
inhomogeneities in which relative density contrasts are $\delta\rho /\rho >1$. 
Such 
inhomogeneities were called {\it archioles}. In the other words {\it archioles} 
are a 
formation that represents a replica of the percolational Brownian vacuum 
structure of 
axionic walls bounded by strings an which is fixed in the strongly inhomogeneous 
initial distribution of axionic CDM. The scale of this modulation of density 
distribution exceeds the cosmological horizon because of the presence of 80\% 
infinite 
component in the structure of axionic wall bounded by strings system. The 
superweakness of the axion field selfinteraction results in the separation of 
archioles 
and of the vacuum structure of axionic walls -- bounded -- by -- strings. So 
these two 
structures evolve independently. The structure of walls bounded by strings 
disappears 
rapidly due to disintegration into separate fragments and further axion 
emission. The 
structure of archioles remains frozen at the RD stage. On the large scales, the 
structure 
of archioles is an initially nonlinear formation, a Brownian network of quasi -- 
one --
dimensional filaments of dustlike matter with the step
\beq
\label{step}
L^A(t)=\lambda\tilde t
\eeq
(where $\lambda \cong 1$). At the moment of creation $\tilde t$, the linear 
density of 
this quasilinear filamentary formations given by
\beq
\label{6}
\mu_A=\pi^2\bar {\rho_a}\tilde t\delta
\eeq
In accordance with this, the cosmological evolution of archioles in the 
expanding 
Universe is reduced to the extension of lines along only one direction.  

We have studied in \Ref{4} the spectrum of inhomogeneities that the density 
develops in response to the large-scale Brownian modulation of the distribution 
of CO 
of axion field. Density perturbations, associated with Brownian network of 
archioles, 
may be described in the terms of a two -- point autocorrelation function 
\Ref{4}. To 
obtain such autocorrelation function, it is necessary to perform averaging of 
energy 
density of infinite Brownian lines over all lines and over the Winner measure, 
which 
corresponds to the position along of Brownian line (see \Ref{4}). 

The two -- point autocorrelation function in the Fourier representation has the 
form 
\beq
\label{7}
\langle\frac{\delta\rho}{\rho_0}(\vec k) \frac{\delta\rho}{\rho_0}(\vec 
k')\rangle 
=12\rho_A\mu_Ak^{-2}\delta (\vec k+\vec k')\tilde t^{-1}f^{-2}t^4G^2
\eeq
where $\rho_0$ is background density, $f_{MD}=3/(32\pi )$ for dustlike stage, 
$f_{RD}=(6\pi)^{-1}$  for RD stage,  $G$ is the gravitational constant, $\rho_A$ 
is 
the total energy density of  the Brownian lines. The mean- square fluctuation of 
the 
mass is given by
\beq
\label{8}
\left(\frac{\delta M}{M}\right)^2(k,t)=12\rho_A\mu_A\tilde t^{-1}f^{-2}G^2kt^4
\eeq
\section{Cosmological impact of archioles}

Let us consider a region characterized at instant $t$ by a size $l$ and a 
density 
fluctuation $\Delta$. For anisotropy of relic radiation we then obtain
\beq
\label{9}
\frac{\delta T}{T}\cong -\delta\left(\frac{l}{t}\right)^2
\eeq
If $l=t$, we have $|\delta T/T|\cong|\Delta |$; that is, the anisotropy of relic 
radiation 
is equal to the density contrast calculated at the instant when the size of the 
region is 
equal to the size of the horizon (Sachs -- Wolf effect). To estimate the 
quadrupole 
anisotropy that is induced in relic radiation by the structure of archioles, we 
must find 
the amplitude of perturbations on the scale of the modern horizon
\beq
\label{10}
\left(\frac{\delta M}{M}\right)^2=2.1\cdot 10^{-
25}\left(\frac{\F}{10^{10}GeV}\right)^4\left(\frac{t_{RD}}{1s.}\right)^{2/3} 
\left(\frac{t_{pres}}{1s.}\right)^{1/3}(k_{hor}t_{pres})
\eeq
Thus Sachs-Wolf  quadrupole anisotropy of relic radiation induced by archioles 
will 
be
\beq
\label{11}
\frac{\delta T}{T}\cong 2.3\cdot 10^{-6}\left(\frac{\F}{10^{10}GeV}\right)^2
\eeq
According to COBE data (see for examle \Ref{cobe}), 
the measured quadrupole anisotropy of relic radiation is at 
the level of
\beq
\label{12}
  \frac{\delta T}{T}\approx 5\cdot 10^{-6}
\eeq
If we take into account the uncertainties of our consideration such as the 
uncertainties 
in correlation length scale of Brownian network ($\lambda \approx 1\div 13$) and 
in 
temperature dependence of axion mass, we can obtain a constraint on the scale of 
symmetry breaking in the model of an invisible axion
\newcommand{\M}{m_a}
\bear
\label{13}
\F\le 1.5\cdot 10^{10}GeV\div 4\cdot 10^{9}GeV;\qquad
\M\ge 410\mu eV\div 1500\mu eV
\ear

This upper limit for $\F$ is close to the strongest upper limits in 
\Refs{5,6,7}, 
obtained by comparing the density of axions from decays of axionic strings with 
the 
critical density, but has an essentially different character. 

The point is that the density of axions formed in decays of axionic strings 
depends 
critically on the assumption about the spectrum of such axions (see \Refs{5,6}) 
and 
on the model of axion radiation from the strings (see \Ref{7}). For example, 
Davis 
\Ref{5} assumed that radiated axions have a maximum wavelength of $\omega 
(t)\cong t^{-1}$ while Harari and Sikivie \Ref{6} have argued that the motion of 
global strings was overdamped, leading to an axion spectrum emitted from 
infinite 
strings or loops with a flat frequency spectrum $\propto k^{-1}$.  This leads to 
an 
uncertainty factor of $\simeq 100$ in the estimate of the density of axions from 
strings and to 
the corresponding uncertainty in the estimated upper limit on $\F$
\bear
\label{14}
\F\le 2\cdot 10^{10}\varsigma GeV;\qquad
\M\ge 300/\varsigma\mu eV
\ear
Here, $\varsigma =1$ for the spectrum from Davis  \Ref{5}, and $\varsigma\approx 
70$ for 
the spectrum from Harari and Sikivie  \Ref{6}. 

In their treatment of axion radiation from global strings, Battye and Shellard 
\Ref{7} found that the dominant source of axion radiation are string loops 
rather than 
long strings, contrary to what was assumed by Davis \Ref{5}. This leads to the 
estimations
\bear
\label{15}
\F\le 6\cdot 10^{10}GeV\div 1.9\cdot 10^{11}GeV; \qquad
\M\ge 31\mu eV\div 100\mu eV
\ear
Arguments that lead to the constraint \Eq{13} are free from these uncertainties, 
since 
they have a global string decay model -- independent character. 

At the smallest scales, corresponding to the horizon in the period $\tilde t$, 
evolution of archioles just in the beginning of axionic CDM dominancy in the 
Universe (at redshifts $z_{MD}\cong 4\cdot 10^4$) should lead to formation of 
the 
smallest gravitationally bound axionic objects with the minimal mass 
$M\simeq \rho_a \tilde t^3\simeq 10^{-6}M_{\odot}$ and of typical minimal size 
$\tilde 
t(1+z_A)/(1+z_{MD})\cong 10^{13}cm$. One can expect the mass distribution of 
axionic objects at small scale to peak around the minimal mass, so that the 
existence 
of halo objects with the mass ($10^{-6} M_{\odot}\div 10^{-1} M_{\odot}$) and 
size 
$10^{13}\div 10^{15}cm$ is rather probable, what may have interesting 
application 
to the theoretical interpretation of MACHOs microlensing events.

\section{Physical impact of archioles}

The inclusion of obtained restriction into the full cosmoparticle analysis 
provides 
detailed quantitative definition of the cosmological scenario, based on the 
respective 
particle physics model. Consider, for example, a simple variant of gauge theory 
of 
broken family symmetry (TBFS) \Ref{8}, which is based on the standard model of 
electroweak interactions and QCD, supplemented by spontaneously broken local 
$SU(3)_H$ symmetry for quark--lepton families. This theory provides natural 
inclusion of Peccei--Quinn symmetry $U(1)_H\equiv U(1)_{PQ}$, being associated 
with heavy "horizontal" Higgs fields and it gives natural solution for QCD CP --
violation problem. The global $U(1)_H$ symmetry breaking results in the 
existence 
of axion--like Goldstone boson -- $a$. TBFS turns to be a simplest version of the 
unified theoretical physical quantitative description of all main types of dark 
matter 
(HDM--massive neutrinos, axionic CDM and UDM in the form of unstable neutrinos 
\Refs{8,9}) and the dominant form of the dark matter is basically determined by 
the 
scale of the "horizontal" symmetry breaking $V_H$, being the new fundamental 
energy scale of the particle theory. For given value of  $V_H$ the model defines 
the 
relative contribution of hot, cold and unstable dark matter into the total 
density.
Since in the TBFS the scale of horizontal symmetry breaking $V_H$ is associated 
with $\F$, we have, from \Eq{13}, the same upper limit on $V_H$. However, this 
limit assumes, that the considered inflationary model permits topological 
defects and 
hence archioles formation due to the sufficiently high reheating temperature 
$T_{RH}\ge V_H$. In the inflationary model, which occurs in TBFS, we can achieve 
$T_{RH}\sim 10^{10}GeV$. 
The "horizontal" phase transitions on inflationary stage lead to the appearance 
of a 
characteristic spikes in the spectrum of initial density perturbations. These 
spike--like 
perturbations, on scales that cross the horizon (60--1) $e$-- folds before the 
end of 
inflation reenter the horizon during the radiation or dust like era and could in 
principle 
collapse to form primordial black holes. The minimal interaction of "horizontal" 
scalars of TBFS $\xi^{(0)}$, $\xi^{(1)}$, $\xi^{(2)}$ with inflaton allows us to 
include them in the effective inflationary potential \Ref{10}:
\bear
\label{16}
V(\phi , \xi^{(0)},\xi^{(1)},\xi^{(2)})=-
\frac{m^2_{\phi}}{2}\phi^2+\frac{\lambda_{\phi}}{4}\phi^4-
\sum_{i=0}^2\frac{m^2_i}{2}\left(\xi^{(i)}\right)^2
+\sum_{i=0}^2\frac{\lambda^{(i)}_{\xi}}{4}\left(\xi^{(i)}\right)^4+\sum_{i=0}^2
\frac
{\nu^2_{\xi}}{2}\phi^2\left(\xi^{(i)}\right)^2
\ear
The analysis of processes of primordial black holes formation from density 
fluctuations, which can be generated by "horizontal" phase transitions at the 
inflationary stage gives rise to an upper limit on the scale of horizontal 
symmetry 
breaking \Ref{10}.
\beq
\label{17}
V_H\le 1.4\cdot10^{13}GeV
\eeq

  Therefore the range between the two upper limit \Eq{13} and \Eq{17} turns to 
be 
not closed, and the following values seem to be possible
\beq
\label{18}
10^{11}GeV\le V_H\le 10^{13}GeV
\eeq
The indicated range corresponds to the case when all the horizontal phase 
transitions 
take place on the dust like stage and $\phi_{c_2}\ll m_{Pl}$. In this case the 
inflationary field $\phi$ oscillates with initial amplitude $\sim m_{Pl}$. 
According to 
\Refs{11,10} it means that any time the amplitude of the field becomes smaller 
then 
$\phi_{c_2}\ll m_{Pl}$, the last  (axion $\xi^{(2)}$) phase transition with 
symmetry 
breaking occurs, and topological defects are produced. Then the amplitude of the 
oscillating field $\phi$ becomes greater than $\phi_{c_2}$, and the 
symmetry is restored again. However, this regime does not continue too long. 
Within 
a few oscillations, quantum fluctuations of the field $\xi^{(2)}$ will be 
generated 
with the dispersion $\langle\left(\xi^{(2)}\right)^2\rangle\simeq\nu^{-
1}_{\xi}\lambda^{1/2}_{\phi}\ln^{-2}1/\nu^2_{\xi}$ \Ref{11}. For 
\beq
\label{19}
m^2_2\le\nu^{-1}_{\xi}\lambda^{1/2}_{\phi}\lambda_{\xi}m^2_{Pl}\ln^{-
2}1/\nu^2_{\xi}
\eeq
these fluctuations will keep the symmetry restored. The symmetry breaking will 
be 
finally completed when $\langle\left(\xi^{(2)}\right)^2\rangle$  will become 
small 
enough. Thus such phase transition leads to formation of topological defects and 
archioles without any need for high -- temperature effects. Substituting the 
typical 
values for potential \Eq{16} such as $m^2_2\approx 10^{-3}V^2_H$, 
$\lambda_{\xi}\simeq 10^{-3}$, $\nu_{\xi}\simeq 10^{-10}$ (see \Ref{10}) we will 
obtain that the condition \Eq{19} means that for the scales 
\beq
\label{22}
V_H\le 10^{-3}m_{Pl}
\eeq
the phenomenon of non -- thermal symmetry restoration takes place in simplest 
inflationary scenario based on TBFS. Owing to this phenomenon oscillations of 
the 
field $\xi^{(2)}$ do not suppress the topological defects and archioles 
production for 
the range \Eq{18}. So the range \Eq{18} turns to be closed by comparison of BBBR 
quadrupole anisotropy, induced by archioles, with the COBE data. As a result, 
the 
upper limit on the scale of horizontal symmetry breaking will be given by 
\Eq{13}. 

Note, in conclusion, that the axion emission can influence the time scale and 
energetics of neutrino flux from collapsing stars. Analysis of this effect for 
SN1987A 
excludes the interval $3\cdot 10^6GeV\le V_H\le 3\cdot 10^9GeV$ (see \Ref{12}) 
and 
establishes the lower limit on the high energetic branch of  TBFS. Thus putting 
together all these limits we can extract narrow window in the high energetic 
branch of 
the so called model of horizontal unification (MHU) \Ref{13}:
\beq
\label{21}
3\cdot 10^9GeV\le V_H\le 1.5\cdot 10^{10}GeV
\eeq

On the base of this choice for the main parameter of MHU we can  build a 
quantitatively definite dark matter scenario, which associates the formation of 
the 
large--scale structure in the Universe with a mixture of axions and massive 
neutrinos, 
since in this interval the total density equal to the critical one makes in the 
most cases 
the contribution of massive neutrinos necessary.

\end{document}